\documentstyle[epsf,epsfig,graphicx]{elsart}
\begin{document}
\begin{frontmatter}
\newcommand{\postscript}[2]{\setlength{\epsfxsize}{#2\hsize}
\centerline{\epsfbox{#1}}}
\title{Trajectory of neutron$-$neutron$-^{18}C$ excited three-body state}
\author[itapeva]{M. T.  Yamashita}, 
\author[ITA]{T. Frederico} and 
\author[IFT]{Lauro Tomio}\footnote{Email: tomio@ift.unesp.br}
\address[itapeva]{Universidade Estadual Paulista, 18409-010, Itapeva, SP, Brazil.}
\address[ITA]{Departamento de F\'\i sica, ITA, CTA, 12228-900 S\~ao Jos\'e dos Campos, Brazil.} 
\address[IFT]{Instituto de F\'\i sica Te\'orica, UNESP,
01405-900 S\~{a}o Paulo, Brazil.}
\date{\today}
\maketitle
\begin{abstract}
The trajectory of the first excited Efimov state is investigated by using a renormalized
zero-range three-body model for a system with two bound and one virtual two-body subsystems.
The approach is applied to $n-n-^{18}$C, where the $n-n$ virtual energy and the three-body
ground state are kept fixed. It is shown that such three-body excited state goes from a
bound to a virtual state when the $n-^{18}$C binding energy is increased.
Results obtained for the $n-^{19}$C elastic cross-section at low energies
also show dominance of an $S-$matrix pole corresponding to a bound or virtual Efimov
state. It is also presented a brief discussion of these findings in the context of 
ultracold atom physics with tunable scattering lengths.
\newline\newline
PACS {03.65.Ge, 21.45.+v, 11.80.Jy, 21.10.Dr}
\end{abstract}
\begin{keyword}
Bound states, scattering theory, Faddeev equation, Few-body
\end{keyword}
\end{frontmatter}

The interest on three-body phenomena occurring for large two-body scattering lengths
have increased in the last years in view of the experimental possibilities presented
in ultracold atomic systems, where the two-body interaction can be tunned by using
Feshbach resonance techniques. Theoretical predictions, well investigated for three particle
systems, such as the increasing number of three-body bound states when the two-body
scattering length goes to infinity - known as Efimov effect~\cite{Efimov} - can actually be
checked experimentally in ultracold atomic laboratories. Indeed, the first indirect evidence of
Efimov states came from recent experiments with ultracold trapped Caesium atoms made by
the Innsbruck group~\cite{grimm}.

In the nuclear context, the investigations of Efimov states are being of renewed
interest with the studies on the properties of exotic nuclei systems with two halo neutrons
($n-n$) and a core ($c$). One of the most promising candidates to present these states is
the $^{20}$C~\cite{amorim,ArPRC04,rmpjensen} ($c\equiv ^{18}C$). $^{20}$C has a ground
state energy of 3.5 MeV with a sizable error in $n-^{18}$C two-body energy,
160$\pm$110 keV~\cite{audi}.

The proximity of an Efimov state (bound or virtual) to the neutron-core elastic scattering cut makes
the cross-section extremely sensitive to the corresponding $S$-matrix pole. However, one should be
aware that the analytic properties of the $S$-matrix for Borromean
systems (where all the two-body subsystems are unbound, like the Caesium atoms of the
Innsbruck experiment~\cite{grimm}) are expected to be quite different from systems where
at least one of the two-body subsystems are bound, as the present case of $n-n-^{18}$C.

The trajectory of Efimov states for three particles with equal masses has been studied
in \cite{adhprc821,adhprc82} using the Amado model~\cite{amado}. By studying the $S-$matrix in
the complex plane, varying the two-body binding, it was confirmed previous analysis
\cite{adhprc82,amadonoble}, that Efimov bound states disappear into the unphysical energy sheet
associated to the unitarity cut, becoming virtual states. It was also verified that,
by further increasing the two-body binding, the corresponding pole trajectories remain
in the imaginary axis and never become resonant.

Considering general halo-nuclei systems ($n-n-c$), in \cite{amorim} it was mapped a
parametric region defined by the $s-$wave two-body (bound or virtual) energies, where the
Efimov bound states can exist.
By increasing the binding energy of a two-body subsystem, it was noted that the
three-body bound state turns out to a virtual state, remaining as virtual
with further increasing of the two-body binding, as already verified in
\cite{adhprc82,virtual}. In the other side, starting with zero two-body binding,
by increasing the two-body virtual energy we have the three-body energy going
from a bound to a resonant state~\cite{resonance}.

Actually, the analysis of the trajectory of Efimov states in the complex plane can be
relevant to study properties of the $^{20}$C.
In order to study the behavior of Efimov states, the authors of \cite{ArPRC04} have recently
pointed out the importance of the analysis of low energy $n-^{19}$C elastic scattering
observables. 
Their results, leading to a $^{20}$C resonance prediction near the scattering
threshold~\cite{ArPRC04}, when the separation energy of the bound halo neutron of 
$^{19}$C is changed, suggest a different behavior from the one found for three
equal-mass particles, where a bound Efimov state turns into a virtual one as the two-body
binding increases. In view of the importance of the results, not only in the nuclear context,
it is interesting to consider a new independent analysis of three-particle systems where 
two subsystems are bound and one is unbound, considering particularly the case 
with different masses. As we show in the present letter, the results of our treatment 
for three-body system with unequal masses are consistent with the expectation derived 
from equal mass systems. Moreover, they are model independent due to the universal character of 
the three-body physics at very low energies. See Ref.~\cite{prlcomment} for a comment 
on the results obtained in Ref.~\cite{ArPRC04}.

In order to clarify the behavior (in the complex energy plane) of
a given Efimov state for the $n-n-^{18}$C system, we consider the
$n-c$ subsystem bound with varying energies; the virtual energy of
the $n-n$ subsystem and the ground state energy of $^{20}$C are
fixed, respectively, at $-$143 keV~\cite{raios,samba} and $-$3.5 keV
(these constraints are convenient to follow the trajetory of the excited
Efimov states since their appearance/disappearance depends only on the
ratio of the $n-c$ and $n-n-c$ energies, see pages 327-329 of Ref.~\cite{braaten}).
 Our present study of $n-n-^{18}$C can be easily extended to similar systems
such as $^{12}$Be, $^{15}$B, $^{23}$N and $^{27}$F.

\begin{figure}[tbh!]
\vspace{-0.2cm}
\centerline{\epsfig{figure=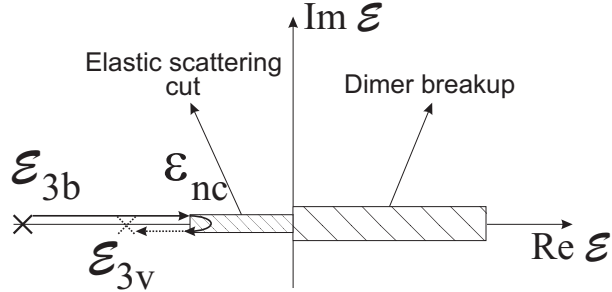,width=8cm}}
\caption{Analytic structure of the $S-$matrix. The energies of the
neutron-core (bound system), the three-body bound and the three-body
virtual states are, respectively, given by $\varepsilon_{nc}$,
${\cal E}_{3b}$, and ${\cal E}_{3v}$.
The arrow passing through the elastic scattering cut shows the trajectory of 
an $S-$matrix pole to the second Riemann sheet.}
\label{cut}
\vspace{-0.3cm}
\end{figure}

In the present case, the bound state equations are extended to the
second Riemann energy sheet through the $n-^{19}$C elastic
scattering cut (see Fig. \ref{cut}) using a well-known technique~\cite{glockle}.
Consistent with previous results~\cite{amorim,adhprc82,virtual} for
three equal masses particles,  by using the present approach that will
be explained in the following, we conclude that
({\it also in this case}) the three-body bound state turns into a virtual
(and not a resonance) one when we increase the $n-^{18}$C two-body binding energy.
The full behavior of the Efimov virtual states in the unphysical sheet of the
complex energy plane, by further increasing the two-body bound state
energy, is still open for investigation.

In our present investigation, we also show that, when the
three-body $S-$matrix pole of a virtual or excited Efimov state is
near the scattering threshold, the $n-^{19}$C elastic scattering
cross section is dominated by such pole, peaking at zero relative
energy and decreasing monotonically with energy.

Next, we introduce the basic formalism by starting  with the coupled
spectator functions for a bound three-body system $n-n-c$ ($c\equiv$ $^{18}C$ in our
specific case). Our units are such that $\hbar=m_n=1$, where $m_n$ is the mass of the
neutron, with the $n-n$, $n-c$ and three-body energies respectively given
by $E_{nn}=\hbar^2\varepsilon_{nn}/m_n$,
$E_{nc}=\hbar^2\varepsilon_{nc}/m_n$, and $E_{3b}=\hbar^2{\cal E}_{3b}/m_n$, where $n-n$
and $n-c$ refer to virtual and bound subsystems. After partial wave projection, the
$\ell$-wave spectator functions $\chi_{n}^\ell$ and $\chi_{c}^\ell$ (where the 
subindex $n$ or $c$ indicates the spectator particle), for a bound three-body 
system, are given by~\cite{amorim}
\vspace{-1cm}{\small
\begin{eqnarray}
{\hspace{-1cm}}\chi_{n}^\ell(q;{\cal E}_{3b})
&=&\tau_{nc}(q;{\cal E}_{3b}) \int_0^\infty \hspace{-0.1cm}dk k^2
    \left[K_2^\ell(q,k;{\cal E}_{3b}) \chi_n^\ell({k};{\cal E}_{3b})
+ K_1^\ell(q,k;{\cal E}_{3b}) \chi_c^\ell({k};{\cal E}_{3b})
   \right], \label{chin} \\
{\hspace{-0.5cm}}\chi_{c}^\ell({q};{\cal E}_{3b})&=&\tau_{nn}(q;{\cal E}_{3b}) \int_0^\infty \hspace{-0.1cm}dk k^2\;
     K_1^\ell(k,q;{\cal E}_{3b}) \chi_n^\ell({k};{\cal E}_{3b}), \label{chic} 
\vspace{-.5cm}\end{eqnarray}
}
where
\vspace{-.5cm}
\begin{eqnarray}
K_{i=1,2}^\ell(q,k;{\cal E}_{3b})&\equiv& G_i^\ell(q,k;{\cal E}_{3b})-\delta_{\ell 0}\;
G_i^\ell(q,k;-\mu^2)
\label{ki} \\
\label{g1}
G_i^\ell(q,k;{\cal E}_{3b})&=&\int_{-1}^{1}\hspace{-0.1cm}dy\frac{P_\ell(y)}
{{\cal E}_{3b}-\frac{A+1}{A+A^{i-1}}q^2- \frac{A+1}{2A}k^2
-\frac{k q y}{A^{i-1}}},
\\
\tau_{nn}(q;{\cal E}_{3b}) &\equiv& \frac{-2}{\pi}\left[
\sqrt{|\varepsilon_{nn}|}+\sqrt{ \frac{A+2} {4A}q^2-{\cal E}_{3b}}
\right]^{-1}, \label{taunn} 
\\
\tau_{nc}(q;{\cal E}_{3b})&\equiv& 
\frac{1}{\pi}\left[\frac{A+1}{2A}\right]^{3/2}
\left[\sqrt{|\varepsilon_{nc}| } -\sqrt{ \frac{
(A+2)}{2(A+1)}q^2-{\cal E}_{3b}}\right]^{-1}\;
. \label{taunc}
\end{eqnarray}
 In the above, 
${\cal E}_{3b}\equiv \varepsilon_{nc}-\frac{A+2}{2(A+1)} \kappa_b^2$
and $A$ is the mass-number of particle $c$. 
The absolute value of the momentum of the spectator particle with 
respect to the center-of-mass (CM) of the other two particles is given
by $q\equiv|\vec{q}|$; with $k\equiv|\vec{k}|$ being the absolute 
value of the relative momentum of these two particles.
The $n-n$ virtual state energy is fixed at $E_{nn}=
-$143 keV. In Eqs. (\ref{chin}) and (\ref{chic}), we have the
Kronecker delta $\delta_{\ell 0}$ ($= 1$ for $\ell=0$ and =0 for
$\ell\ne 0$) in order to renormalize the equations (using a
subtraction procedure) only for the partial wave where such
renormalization is necessary, $\ell=0$. In the cases of $\ell>0$,
due to the centrifugal barrier, the Thomas collapse is absent and
such renormalization is not necessary. In this way, with Eqs.
(\ref{chin}) and (\ref{chic}) renormalized, the three-body
observables are completely defined by the two-body energy scales,
$\varepsilon_{nc}$ and $\varepsilon_{nn}$.  Later on, the
definitions (\ref{ki})-(\ref{taunc}) will be extended also to
unbound systems.
The regularization scale $\mu^2$, used in the $s-$wave [see Eq.~(\ref{ki})], 
is chosen to reproduce the three-body ground-state energy of $^{20}$C,
$E_0=-$3.5 MeV~\cite{audi}.
A limit cycle~\cite{braaten,mohr} for the scaling function of $s-$wave observables
is evidenced when $\mu$ is let to be infinity. We note that, a good description of 
this limit is already reached in the first cycle~\cite{virtual}.

The analytic continuation of the bound three-body system given
by (\ref{chin}) and (\ref{chic}) to the second Riemann sheet is 
performed through the $n-(n-c)$ elastic
scattering cut, following Refs.~\cite{virtual,glockle,frederico}.
As we are considering $n-n-c$ nuclei where only the subsystem $n-c$
is bound (``samba-type" nuclei~\cite{raios,samba}), we have only
the $n-c$ cut in the complex energy plane. 
In order to see how to perform the analytical continuation 
from the first to the second sheet of the complex energy, let us first 
consider a complex momentum variable $k_i$. As the energies of the two-body
sub-systems are fixed and only the $n-c$ subsystem is bound, it is convenient 
to define this momentum as  $k_i=\sqrt{\frac{2(A+1)}{A+2}
\left({\cal E}_i-\varepsilon_{nc}\right)}$. In this case, the bound-state energy 
$({\cal E}_i={\cal E}_{3b})$ is given by $k_i={\rm i}\kappa_b$, with 
the virtual state energy (${\cal E}_i={\cal E}_{3v}$) given by 
$k_i=-{\rm i}\kappa_v$.
Next, by removing ${\cal E}_{3b}$ in favor of $k_i$ in the bound-state equations 
(\ref{chin})-(\ref{taunc}), with 
{\small \begin{equation}
{\bar\tau_{nc}}(q;{\cal E})\equiv
\frac{-1}{\pi}
\left(\frac{A+1}{2A}\right)^{\frac 32}
\hspace{-0.1cm}\left(\hspace{-0.2cm}\sqrt{|\varepsilon_{nc}}|+\hspace{-0.1cm}\sqrt{
  \frac{(A+2)q^2}{2(A+1)}-{\cal E}}\right),
\label{tauncbar}
\end{equation}}\vspace{-.5cm}
and with $\chi_c$ and $\chi_n$ redefined as
\begin{equation} 
\chi_c^\ell({q};{\cal E})\equiv h_c^\ell({q};{\cal E}),\;\;
\displaystyle \chi_n^\ell({q};{\cal E})\equiv
h_n^\ell({q};{\cal E})/ 
{(q^2-k_i^2-i\epsilon)},
\label{hchn}\end{equation}\vspace{-.4cm}
we observe that the relevant integrals to be considered have the structure 
${\cal I}(k_i) = \int q^2 dq \left[{F(q^2)}/{(k_i^2-q^2+{\rm i}\epsilon)}\right]$
in the first sheet of the complex energy plane. To obtain ${\cal I}(k_i)$ in the 
second sheet we need to change the contour of integration, as shown in detail in 
Ref.~\cite{adhprc821}, such that the value of  ${\cal I}(k_i)$ in the second sheet 
of the complex energy plane is given by 
${\cal I^\prime}(k_i) = {\cal I}(k_i) - {\rm i }\pi k_i F(k_i^2).$
Following this procedure, from Eqs. (\ref{chin}) and (\ref{chic}) we obtain
the corresponding equations in the second sheet of the complex energy plane, where
for the virtual state energy we have $k_i=-{\rm i}\kappa_v=\sqrt{\frac{2(A+1)}{A+2}
\left({\cal E}_{3v}-\varepsilon_{nc}\right)}$:
{\small
\begin{eqnarray}
\label{hn}
{\hspace{-.5cm}}h_n^\ell(q;{\cal E}_{3v})=&&\frac{2(A+1)}{A+2}{\bar\tau_{nc}}(q;{\cal E}_{3v})
\left[{\pi}{\kappa_v}K_2^\ell(q,-{\rm i}\kappa_v;{\cal E}_{3v}) h_n^\ell(-{\rm i}\kappa_v;{\cal E}_{3v})
+\right.\nonumber\\ && +\left. \int_0^\infty {\hskip -0.1cm}dk k^2
{\hspace{-0.1cm}}\left({\hspace{-0.1cm}}K_1^\ell(q,k;{\cal E}_{3v}) h_c^\ell(k;{\cal E}_{3v})
+ \frac{K_2^\ell(q,k;{\cal E}_{3v})\;h_n^\ell(k;{\cal E}_{3v})}{k^2+\kappa_v^2}
{\hspace{-0.1cm}}\right){\hspace{-0.1cm}}\right],
\\ \label{hc}
{\hspace{-.5cm}}h_c^\ell(q;{\cal E}_{3v})=&&{\hspace{-0.1cm}}\tau_{nn}(q;{\cal E}_{3v})
\left[{\pi}{\kappa_v}  K_1^\ell(-{\rm i}\kappa_v,q;{\cal E}_{3v}) h_n^\ell(-{\rm i}\kappa_v;{\cal E}_{3v})
+\right.\nonumber\\ &&+\left. \int_0^\infty \hspace{-0.1cm}dk k^2\;
\frac{ K_1^\ell(k,q;{\cal E}_{3v})\;h_n^\ell(k;{\cal E}_{3v})}{k^2+\kappa_v^2}\right].
\end{eqnarray}}
The above coupled equations, as well as the corresponding coupled equations for bound state, 
can be written as single-channel equations for $h_n^\ell$, by defining an effective interaction ${\cal V}$ and considering $I=b,v$:
\vspace{-0.5cm}{\small
\begin{eqnarray}
\label{hn2}
{\hspace{-.8cm}}
h_n^\ell(q;{\cal E}_{3I})&=&
2{\kappa_v} h_n^\ell(-{\rm i}\kappa_v;{\cal E}_{3v})
{\cal V}^\ell(q,-{\rm i}\kappa_v;{\cal E}_{3v})\delta_{I,v}+
\frac{2}{\pi}\int_0^\infty {\hskip -0.1cm}dk k^2
{\cal V}^\ell(q,k;{\cal E}_{3I})
\frac{h_n^\ell(k;{\cal E}_{3I})}{k^2+\kappa_I^2}
\nonumber\\
\end{eqnarray}
\vspace{-1.5cm}
\begin{eqnarray}
{\hspace{-.5cm}}{\cal V}^\ell(q,k;{\cal E}_{3I})&\equiv&
\pi\frac{(A+1)}{A+2}{\bar\tau_{nc}}(q;{\cal E}_{3I})
\times \\ \times &&
\left[K_2^\ell(q,k;{\cal E}_{3I})+
\int_0^\infty \hspace{-0.1cm}dk'k'^2
K_1^\ell(q,k';{\cal E}_{3I})\tau_{nn}(k';{\cal E}_{3I})
K_1^\ell(k,k';{\cal E}_{3I})
\right]. 
\nonumber \label{V}
\end{eqnarray}
}
The first term in the right-hand-side (rhs) of (\ref{hn2}), 
 which is non-zero only for the virtual state, corresponds
to the contribution of the residue at the pole.
The virtual states are limited by the cut of the elastic scattering
amplitude in the complex plane, corresponding to the second term
in the rhs of Eq.~(\ref{hn}).
In this case, the cut is given by the zero of the
denominator of $G_2^\ell(q,k;{\cal E}_{3v})$ [See Eq. (\ref{g1})], where
$-1 < y < 1$ and $q=k=-{\rm i}\kappa_{cut}$.
With $|{\cal E}_{3v}|=|\varepsilon_{nc}|+\frac{A+2}{2(A+1)}\kappa_{cut}^2$,
we obtain the branch points, with the cut given by
\begin{equation}
\frac{2(A+1)}{A+2}|\varepsilon_{nc}|<|{\cal E}_{3v}|<\frac{2(A+1)}{A}|\varepsilon_{nc}|. \label{cut1}
\end{equation}
For $n-n-^{18}$C ($A=18$), a virtual state
energy can be found in the energy interval between the threshold of the elastic
scattering and the starting of the above cut (\ref{cut1}):
\begin{equation}
|\varepsilon_{nc}|\;<\;|{\cal E}_{3v}|\;<\;1.9\;|\varepsilon_{nc}| \ .
\end{equation}
\begin{figure}[thb!]
\centerline{\epsfig{figure=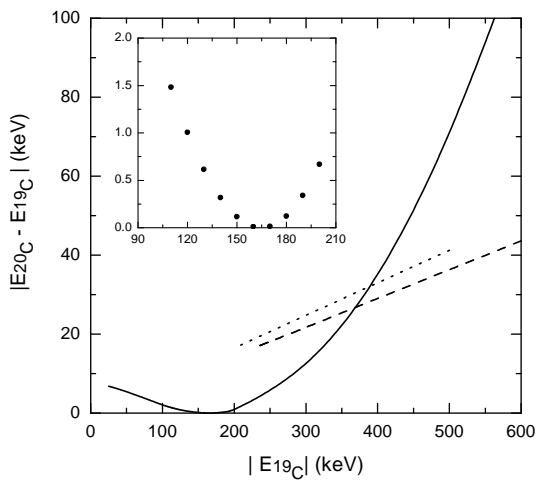,width=7cm}}
\vspace{-0.3cm}\caption{
Three-body $n-n-^{18}$C results for the first excited state, with respect
to the threshold ($|E_{^{20}{\rm C}}-E_{^{19}{\rm C}}|$) for varying
$^{19}$C binding energies. Three-body bound (virtual) states occur when
$|E_{^{19}{\rm C}}|$ is approximately smaller (larger) than 170 keV.
$s-$wave results (solid line) are also presented in the inset (with dots).
Results for the $p-$and $d-$waves, divided by a factor 10, are shown with
dashed and dotted lines, respectively.}
\label{virtual}
\end{figure}
Figure \ref{virtual} shows how the absolute value of the first excited three-body state
energy with respect to the two-body bound
state, $|E-E_{nc}|$, varies when increasing the $n-c$ bound-state energy.
In this case, with $^{18}$C being the core, we have $^{19}$C
as the two-body bound subsystem.
The solid line of Fig. \ref{virtual} presents the $s-$wave results, with a close
focus in the region of the threshold ($E=E_{nc}$) given by the inset figure. 
Although the Efimov states do not appear in higher partial waves due to the existence 
of the centrifugal barrier, we have also presented results for the virtual states of 
$p-$ and $d-$ waves in view of their possible relevance for the elastic $n-^{19}$C 
low-energy cross-section. Such situation can happen when the $s-$wave virtual state is  
far from the elastic scattering region.
\newline\indent
The results shown in Fig.~\ref{virtual}, valid for $n-n-c$ system with
$n-c$ bound, together with previous analysis~\cite{amorim,virtual}, are clarifying
that the behavior of Efimov states,
when one of the particles have a mass different from the other two, follows the same
pattern as found in the
case of three  equal-mass particles~\cite{adhprc82,amadonoble}.
No resonances were found for Eqs. (\ref{hn}) and (\ref{hc}) in the complex energy plane.
\newline\indent
Next, for consistency, we present results for the $s-$wave elastic cross-sections,
which are in agreement with the above.
The formalism for the partial-wave elastic $n-~^{19}$C scattering
equations can be obtained from Eqs.~(\ref{chin}) and
(\ref{chic}), by first introducing the following boundary condition
in the full-wave spectator function $\chi_n(\vec q)$:
\vspace{-0.3cm}
\begin{equation}
\chi_n(\vec q)\equiv(2\pi)^3\delta(\vec q-\vec k_i)
+4\pi\frac{h_n(\vec q;{\cal E}(k_i))}{q^2-k_i^2-{\rm i}\epsilon} ,
\end{equation}\vspace{-0.3cm}
where $h_n(\vec q;{\cal E}(k_i))$ is the scattering amplitude, and
the on-energy-shell incoming and final relative momentum are
related to the three-body energy ${\cal E}_i\equiv{\cal E}(k_i)$ by
$ k_i\equiv |\vec k_i|=|\vec k_f| =
 \sqrt{[2(A+1)/(A+2)]\left({\cal E}_i-\varepsilon_{nc}\right)}.
$
With the same formal expressions (\ref{tauncbar}) and (\ref{taunc})
for $\bar\tau_{nc}$ and $\tau_{nn}$, by using the definition (\ref{V}), 
the partial-wave scattering equation can be cast in the following single 
channel Lippmann-Schwinger-type equation for $h_n^\ell$:
\vspace{-0.3cm}
{\small
\begin{eqnarray}\label{fnsca}
&&\hspace{-.5cm}h_n^\ell(q;{\cal E}_i)={\cal V}^\ell(q,k_i;{\cal E}_i)
+\frac{2}{\pi}\hspace{-0.2cm}\int_0^\infty \hspace{-0.1cm}dk k^2\;
\frac{{\cal V}^\ell(q,k;{\cal E}_i)\;
h_n^\ell(k;{\cal E}_i)}{k^2-k_i^2-{\rm i}\epsilon} .
\end{eqnarray}}
Virtual states and resonances correspond to poles of the scattering matrix on
unphysical sheets. So, the most natural method to look for them is to perform
an analytically continuation of the scattering matrix to the unphysical sheet.
For that one needs prior knowledge about the analytic properties of the kernel
of the integral equation, as well as the scattering matrix properties on the 
unphysical sheet of the complex energy plane. Although such properties are easy 
to derive in simple cases, they can be difficult to obtain in more complex 
situations, which can put some restriction on the approach.

For the numerical treatment of Eq.~(\ref{fnsca}), we consider the
approach developed in \cite{adhprc82} to find virtual states and
resonances on the second energy sheet associated with the lowest
scattering threshold. Such approach does not require prior knowledge
of the analytic properties of the scattering matrix on the
unphysical sheet. The solution of the scattering equation is written
in a form where the analytic structure in energy is clearly
exhibited. Then, it is analytically continued to the unphysical
sheet. The method relies on the calculation of an auxiliary
(resolvent) function~\cite{adhprc81}, which has an integral
structure similar to the original scattering equation, but with a
weaker kernel due to a subtraction procedure at an arbitrary fixed
point $\bar k_i$. For real positive energies, this subtraction point
$\bar k_i$ is identified with $k_i$, such that the corresponding
integral equation does not have the two-body unitarity cut. For
non-real or negative energies, $\bar k_i$ can be any arbitrary
positive real number. For convenience, $\bar k_i = |k_i|$. The final
solution is obtained by evaluating certain integrals over the auxiliary function.
In case of scattering solutions, the two-body unitarity cut is introduced through
these integrals.
In the present case, we have the following integral equation for the
auxiliary function $\Gamma$, and the corresponding solution for
$h_n^\ell(q;{\cal E}_i)$:
\vspace{-0.5cm}
{\small
\begin{eqnarray}
{\Gamma_n^\ell}(q,k_i;{\cal E}_i)&=&{\cal V}^\ell(q,k_i;{\cal E}_i)
+\frac{2}{\pi}\int_0^\infty\hspace{-0.1cm} dk \left[k^2{\cal
V}^\ell(q,k;{\cal E}_i)-\bar k_i^2{\cal V}^\ell(q,k_i;{\cal E}_i)\right]
\frac{{\Gamma_n^\ell}(k,k_i;{\cal E}_i)}{k^2-k_i^2},
\nonumber\\
{h_n^\ell}(q;{\cal E}_i)&=&{\Gamma_n^\ell}(q,k_i;{\cal E}_i)+
{\Gamma_n^\ell}(q,\bar k_i;{\cal E}_i)
\frac{\displaystyle \frac{2}{\pi}{\bar k_i^2}\int_0^\infty
dk \;\frac{{\Gamma_n^\ell}(k,k_i;{\cal E}_i)}{k^2-k_i^2-{\rm i}\epsilon}}
{\displaystyle
1-\frac{2}{\pi}{\bar k_i^2}\int_0^\infty
dk \;\frac{{\Gamma_n^\ell}(k,{\bar k_i};{\cal E}_i)}{k^2-k_i^2-{\rm i}\epsilon}}
.\label{hna2}
\end{eqnarray}
}
For the on-shell scattering amplitude, we have
\begin{eqnarray}
&&h_n^\ell(k_i;{\cal E}_i)=
[1/{\Gamma_n^\ell}(k_i,k_i;{\cal E}_i)-{\cal J}]^{-1}
\label{a3}\\
&&{\cal J}\equiv
\frac{2}{\pi}k_i^2\int_0^\infty
dk \;\frac{\left[{\Gamma_n^\ell}(k,k_i;{\cal E}_i)/
{\Gamma_n^\ell}(k_i,k_i;{\cal E}_i)-1\right]}{k^2-k_i^2}
+ {\rm i}k_i
\nonumber 
\end{eqnarray}

In order to obtain the bound and virtual energy states, two
independent procedures have been used. The first one, by solving
directly the homogeneous coupled equations (\ref{chin}),
({\ref{chic}}), (\ref{hn}) and (\ref{hc}), looking for zeros of the
corresponding determinants. The other one, by verifying the position
of the poles in the complex energy plane of the scattering
amplitude, $h_n^\ell(k_i;{\cal E}_i)$, given by Eq.~(\ref{a3}) and
corresponding analytic extension to the second Riemann sheet. In
this second approach, we solve the corresponding inhomogeneous
equation with on-energy-shell momentum $k_i=+{\rm i}\kappa_b$ (bound
state, ${\cal E}_i= -|{\cal E}_{3b}|$) and $k_i=-{\rm i}\kappa_v$
(virtual state, ${\cal E}_i= -|{\cal E}_{3v}|$)\cite{adhprc82}.

By comparing the results of both approaches, we checked that they
give consistent results. However, for numerical stability and
accuracy of the results, particularly in the case of the search for
Efimov states, when the absolute values of the energies are very
close to zero, the second approach is by far much better.

\begin{figure}[tbh!]
\centerline{\epsfig{figure=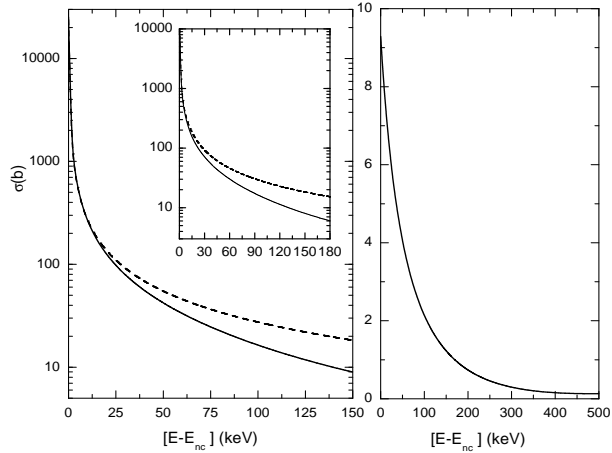,width=8cm,height=6cm}}
\caption{$n-^{19}$C elastic cross sections (in barns)
versus the CM kinetic energies [Eq.~(\ref{ke}) with A=18], for different
${^{19}}$C bound energies.
In the left-hand-side frame we show results for two cases that generate three-body
energies close to the threshold:
$E_{^{19}{\rm C}}=-$150 keV (main figure) and $-$180 keV (inset),
producing respectively three-body bound and virtual states.
Solid-line are obtained from (\ref{fnsca}), with dashed-line from (\ref{sigeff}).
In the rhs we have results for $E_{^{19}{\rm C}}=-$500 keV.
}
\label{xsec}
\end{figure}

Results for the total $n-^{19}$C elastic cross-sections,
obtained from ${d\sigma}/{d\Omega}=|h_n(\vec k_f;{\cal E}_i)|^2$,
refering to bound or virtual states $E^*$, are presented in Fig.~\ref{xsec}
as functions of the CM kinetic energy,
\begin{equation}
{\cal K}(k_i)\equiv {[(A+2)\;k_i^2]}/{[2(A+1)]}
= E(k_i) - E_{nc}.
\label{ke}\end{equation}
Although each higher partial $\ell-$wave have a virtual state, below the
breakup the cross-section is completely dominated by the $s-$wave~\cite{frederico}.
In the frame shown in the left-hand-side, we have two cases of energies close to the 
scattering threshold: $n-{^{18}{\rm C}}$ bound at $-$150 keV, giving an excited Efimov
bound state with $E^*=-$150.12 keV; and $n-{^{18}{\rm C}}$ bound at $-$180 keV,
producing a {\it virtual state} with $E^*=-$180.12 keV. In both two cases the cross-section 
has a huge peak at zero energy due to the presence of the nearby pole.
For comparison, we also show (with dashed-line) the results obtained from the following 
effective range expansion (approximately valid for small $k_i$ near the elastic scattering threshold):
{\small
\begin{eqnarray}
\sigma(k_i) =\frac{4\pi \hbar^2}
{\displaystyle
{1.9 m_n}
\left(E_{nc}-E^*\right)+\hbar^2 k_i^2}
\simeq
\frac{\displaystyle
2741\;{\rm keV}}
{E(k_i)-E^*}\; {\rm barn}, \label{sigeff}
\end{eqnarray}}
where we have used $A=$18, $\hbar^2 / m_n =$414.42 keV barn and Eq.~(\ref{ke}).
The case not so close to the threshold (where Eq.(\ref{sigeff}) fails) is shown
in rhs for $E_{nc}= -$500 keV, with $^{20}$C virtual energy  $E^*=-$568.73 keV.
\newline\indent
The proximity of an Efimov state (bound or virtual) to the neutron and neutron-core
elastic scattering makes the cross-section extremely sensitive to the corresponding
$S$-matrix pole. We remark that if it will be possible to dissociate a ``samba-type"
halo nuclei, like $^{20}$C, and measure the correlation function in the two-body channel
corresponding to a neutron and a bound $n-c$ system for small relative momentum, the
information on the final state interaction as well as the halo structure will be clearly
probed, as the counterpart seen in the $n-n$ correlation in the
breakup of Borromean nuclei~\cite{nn}.

In conclusion, we analyzed the three-body halo system $n-n-^{18}$C, where two pairs 
($n-^{18}$C) are bound, and the remaining pair $n-n$ has a virtual-state.
We study the trajectory of three-body Efimov states in the complex energy plane. 
As shown, the energy of an excited  Efimov state varies from a bound to a virtual 
state as the binding energy of the subsystem $n-^{18}$C is increased, while keeping fixed
the $^{20}$C ground-state energy and the virtual energy of the remaining pair ($n-n$).
In our approach we applied a renormalized zero-range model, valid in the limit of
large scattering lengths. Considering that low-energy correlations, as the one represented 
by the Phillips line~\cite{phillips} (correlation between triton and doublet neutron-deuteron 
scattering length), are well reproduced by zero-range potentials~\cite{zrxphil}, the present 
conclusions should remain valid also for finite two-body interactions when the scattering 
length is much larger than the potential range.
On the numerical analysis, we should remark that we have considered two approaches that 
give consistent results for bound and virtual state energies. 
In the case of Efimov physics, where the poles are very close to zero, the method
considered in Ref.~\cite{adhprc82} was found to give solutions with much better 
stability and accuracy in the scattering region than by using a contour deformation 
technique.

The present results are extending to $n-n-$core systems the long ago conclusion
reached for three equal-mass particles~\cite{adhprc82,amadonoble}:
by  increasing the binding energy of the $n-$core subsystem, an excited
weakly-bound three-body Efimov state moves to a virtual one and will not
become a resonance.  From Fig.~\ref{xsec}, one can also observe that the $n-^{19}$C 
elastic cross-sections at low energies present a smooth behavior dominated by the $S-$matrix 
pole corresponding to a bound or virtual three-body state. In contrast with the above 
conclusion applied to system where $n-c$ is bound, we should observe that it was also 
verified that an excited Efimov state can go from a bound to a resonant state (instead of virtual 
state) in case of Borromean systems (with all the subsystems unbound), when the absolute value of 
a {\it virtual-state energy} for the $n-c$ system is increased~\cite{resonance}. Actually, it 
should be of interest to extend the present analysis of the trajectory of Efimov states to 
other possible two-body configurations, with different mass relations of three-particle systems.

In view of the exciting possibilities of varying the two-body interaction, it can be of high 
interest the results of the present study to analyze properties of three-body systems in ultracold 
atomic experiments. For negative scattering lengths the Efimov state goes to a continumm resonance 
when $|a|$ is decreased, as observed by the change of resonance peak in the three-body 
recombination to deeply bound states towards smaller values of $|a|$ by raising the 
temperature~\cite{pla}. Alternatively, for positive $a$ the recombination rate has a peak 
when the Efimov state crosses the threshold and turns into a virtual state when decreasing $a$.
A dramatic effect will appear in the atom-dimer scattering rate when the cross-section is 
dominated by the $S$-matrix pole near the scattering threshold. 
We foreseen that the coupling between atom and molecular condensate will respond strongly 
to the crossing of the triatomic bound state to a virtual one by changing $a$. Determined 
by the dominance of the coupled channel interaction, new condensate phases of the 
atom-molecule gas are expected. The proximity of the virtual trimer state to the physical region, 
implying in a large negative  atom-dimer scattering length, will warrant stability to both 
condensates, while the positive atom-dimer scattering length, due to a trimer bound state near 
threshold, make possible the collapse of the condensed phases.   Indeed, the occurrence of 
some interesting effects in the condensate due to Efimov states near the scattering threshold 
have already been discussed in Refs.~\cite{newphases}. We hope these exciting new 
consequences of Efimov physics can be explored experimentally in the near future.

LT thanks Prof. S.K. Adhikari for helpful suggestions. We also thank Funda\c c\~ao de Amparo 
\`a Pesquisa do Estado de S\~ao Paulo and Conselho Nacional de Desenvolvimento
Cient\'\i fico e Tecnol\'ogico for partial support.

\end{document}